\documentstyle[12pt]{article}
\def\lapproxeq{\lower .7ex\hbox{$\;\stackrel{\textstyle <}{\sim}\;$}}
\def\gapproxeq{\lower .7ex\hbox{$\;\stackrel{\textstyle >}{\sim}\;$}}

\begin{document}
\titlepage
\begin{flushright}
DTP/94/36 \\
hep-ph/9405328
\end{flushright}

\vspace*{2.5cm}
\begin{center}
{\bf QCD PHYSICS AT LEP 2}
\end{center}

\begin{center}
V.A.\ Khoze\footnote{Presented at the XXIXth Rencontre de Moriond \lq\lq QCD
and
High Energy Hadronic Interactions", M\'{e}ribel, France, March 19-26, 1994.} \\
Department of Physics, University of Durham, Durham DH1 3LE, England
\end{center}

\begin{center}
and
\end{center}

\begin{center}
Institute for Nuclear Physics, 188350, St.\ Petersburg, Gatchina, Russia.
\end{center}

\vspace*{5cm}
\begin{abstract}
I will discuss, in a concise way, the main objectives of QCD studies at LEP 2.
\end{abstract}

\newpage
\noindent  {\bf 1.  Introduction}

I was nominated by the Program Committee of this wonderful conference to play
the role of a scapegoat by covering the prospects for QCD studies at LEP 2.
Focusing on this subject one immediately becomes an easy target for sceptics.
The scepticism is rooted in the unfriendly statistics of hadronic events which
limits the potential of high accuracy measurements at LEP 2.  Recall that with
total integrated luminosity of 500 pb$^{-1}$ per experiment$^{[1,2]}$ the rates
of annihilation events will be reduced considerably, typically by 3 orders of
magnitude, as compared to the $Z^0$ peak. There is also a somewhat
psychological
problem.  Contrary to the exciting prospects for Higgs searches or studies of
New Physics, the QCD program deals with the standard, so-called \lq\lq
down-to-earth" physics.  Moreover, up to now the perturbative QCD scenario has
worked too well, see e.g.\ Refs.\ [3-5].  In some sense QCD is a victim of its
own successes.

I cannot resist mentioning that although LEP 2 will start operating in about
two
years, its physics program until now has not been of primary interest and the
efforts of the theoretical community have been concentrated mainly on the far
more distant NLC, see e.g.\ [6,7].

Now let us move on to the sunny side.  After all, at LEP 2 we enter an
unexplored energy region and there are definite advantages here over the
existing $e^+e^-$ colliders even for the standard QCD studies.
\begin{enumerate}
\item  One would expect a higher credibility for the perturbative predictions
since the subasymptotic corrections become less important.  At the same time,
since multiplication in the QCD cascades grows quite rapidly with increasing
energy ($\sim e^{\sqrt{\ell ns}}$), the nontrivial perturbative
predictions$^{[8,9]}$ should become more spectacular.
\item  Collimation of a jet around the parent parton momentum grows as energy
increases.  Moreover, the collimation of an energy flow grows much more rapidly
compared to a multiplicity flow.  This is of special interest when studying the
dynamics of interjet particle production, see [9] for details.
\item  LEP 2 can provide a remarkable laboratory for testing QCD in
photon-photon reactions.  Both effective energy and luminosity of
$\gamma\gamma$
collisions are here higher than at LEP1.  At the same time the background due
to
annihilation events is incomparably smaller.
\item Some interesting vistas on QCD studies arise in hadronic $W^+W^-$ and
$Z^0\gamma$ events.
\item  LEP 2 results will provide a testing ground for the experiments at the
future NLC facilities.
\item  Recall also the important subsidiary role of the QCD processes.  They
are
always going to give a background to whatever other physics one is interested
in.  Therefore their detailed knowledge is quite mandatory.
\end{enumerate}

There are some auxiliary virtues of LEP 2 for QCD tests.
\begin{enumerate}
\item The operation is expected to start from the $Z^0$ resonance.  This will
provide especially good calibration for many technical and physical purposes.
\item  The $b$-tagging efficiency is expected to be stretched to the limit.
This is driven, first of all, by its crucial role in Higgs identification at
LEP
2.
\end{enumerate}

Finally, recall that the $W^+W^-$, $Z^0\gamma$ processes, which are interesting
in their own right, will complicate a comparison of event characteristics with
QCD.  However, it seems that the appropriate experimental cuts will not
introduce  major additional distortions of the QCD expectations, see e.g.\
[6,7].

\vspace*{1cm}

\noindent  {\bf 2.  Standard QCD Tests}

Hadronic final states of high energy $e^+e^-$ annihilation have traditionally
been a fruitful testing ground for QCD.  We discuss here the standard QCD tests
from studies of the hadronic jet profiles.  We concentrate on the aspects of
the
so-called semisoft and hard QCD physics.

\vspace*{.5cm}

\noindent  2.1.  QCD studies in the semisoft region

During the past years LEP, SLC and TRISTAN have provided an exceedingly rich
source of information on QCD jets, see [3-5].  Moreover, the first experimental
tests of the QCD cascading picture and, in particular, of the spectacular
colour
coherence effects have been successfully performed at the Tevatron$^{[10]}$ and
at HERA$^{[11]}$.   The data demonstrate fairly good agreement with the results
of the analytic perturbative approach (MLLA + LPHD)$^{[8]}$ which has been
developed in the last decade by the international CIS
collaboration\footnote{CIS
$\equiv$ \underline{C}olumbia
University/\underline{C}ambridge-\underline{I}taly-\underline{S}t.\ Petersburg
(the names can be found in [8]).  The name of this collaboration has been
changed following the renaming campaign in the FSU.}.

The well-developed Monte-Carlo programs based on a QCD parton shower mechanism
(the so called WIG'ged MC's\footnote{WIG $\equiv$ \underline{W}ith
\underline{I}nterfering \underline{G}luons (HERWIG, JETSET, ARIADNE).})
describe
the existing data very successfully and provide very useful tools for
experimentalists.  Despite the fact that all these models are of a
probabilistic
and iterative nature, up to now their predictions for the gross jet
characteristics have been in peaceful coexistence with the analytic results.
Unfortunately, LEP 2 does not appear to be the best place to observe the
breakdown of this coexistence (for discussion of the subtle QCD collective
phenomena see refs.\ [8,9]).

Much of the semisoft physics is concerned with distributions and correlations
of
particles in jets in circumstances where colour coherence is important.  The
coherent effects give rise to the so-called hump-backed shape of particle
spectra.  This striking perturbative prediction is very well confirmed
experimentally (see e.g.\ [12]).  The analytic results for particle spectra can
be expressed in terms of confluent hypergeometric functions$^{[8]}$.  The
simplified version, in the case when the cutoff parameter in cascades $Q_0 =
\Lambda$ (the so-called limiting spectrum), is quite convenient for numerical
calculations.  In the context of the LPHD picture the limiting formulae are
applied for the description of $\pi$'s and for all charged particle
spectra$^{[13]}$.  To approximate the distributions of identified massive
hadrons the partonic formulae truncated at the different $Q_0$ values can be
used.  One can encode the MLLA effects in terms of a few analytically
calculated
shape parameters by means of the distorted Gaussian ($\ell = \ell n(1/x_p),
\delta = (\ell - <\ell >)/\sigma$)$^{[13,14]}$:
\begin{equation}
\bar{D}(\ell,Y,\lambda) = \frac{N(Y,\lambda)}{\sigma\sqrt{2\pi}} {\rm exp}
\left[ \frac{1}{8}k - \frac{1}{2} s\delta - \frac{1}{4}(2 + k)\delta^2 +
\frac{1}{6} s\delta^3 + \frac{1}{24} k\delta^4 \right] .
\end{equation}
Here $Y = \ell n(E/Q_0), \lambda = \ell n(Q_0/\Lambda), 2E = W$, the total
c.m.s.\ energy.

An important measure is the position of the maximum of the spectrum,
$\ell_{max}$.  For the limiting case, the energy dependence is
\begin{equation}
\ell_{max} = Y\left[ \frac{1}{2} + \sqrt{\frac{C}{Y}} - \frac{C}{Y} + ...
\right]
\end{equation}
with $C $ = 0.2915(0.3513) for $n_f$ = 3(5).  This is in surprisingly good
agreement with the observed energy evolution of the peak position, see Fig.\ 1.
\begin{figure}
\vspace*{10cm}
\end{figure}
It is predicted that for the truncated distributions the energy dependence of
$\ell_{max}$ is practically universal and the difference in $Q_0$ values leads
only to approximately constant shift of the peak$^{[13]}$.  Measurements at LEP
2 could provide a new insight in studies of the perturbative universality for
different particle species.  An intriguing puzzle, intimately related to the
various aspects of the LPHD concept, is the unravelling of the connection
between $Q_0$ and the particle masses and their quantum numbers.

The energy evolution of particle distributions in the energy region from
PETRA/PEP to NLC is illustrated in Fig.\ 2.
\begin{figure}
\vspace*{10cm}
\end{figure}
It clearly demonstrates the rise of $\ell_{max}$ with increasing $W$ and a fast
growth of the hump height reflecting the rise of jet multiplicity.  The
exploration of LEP 2 energy domain would provide the possibility of further
tests of the underlying cascading dynamics of multiple hadroproduction in jets.

We turn now to studies of the properties of heavy flavour jets.  The analytic
perturbative formulae$^{[15]}$ allow one to predict the inclusive distribution
of a heavy quark $Q$.  They account for all significant logarithmically
enhanced
contributions in high orders including the two-loop anomalous dimension and the
proper coefficient function with exponentiated Sudakov-type logs, as well as
the
controllable dependence on the heavy quark mass $M_Q$.  These results are
expected to describe the energy distributions averaged over heavy-flavour
hadron
states.  Such a situation naturally appears, e.g., when studying inclusive hard
leptons from heavy quark initiated events.  Confronting the perturbative
results
with the data allows one to determine the scale parameter $\Lambda$ and to
quantify the low-momentum behaviour of the effective coupling.  The first
comparison of perturbative predictions with the experimentally measured mean
scaled energies $<x>_{c,b}$ at different $W^{[16]}$ looks rather encouraging.
Further studies in the wide energy region could be quite instructive.  At LEP 2
one can expect about a hundred identified $b \rightarrow$ lepton events and a
precision in measurements of $<x_{b \rightarrow {\rm lepton}}>$ at the 1\%
level$^{[17]}$.

The mean multiplicity in $e^+e^- \rightarrow Q\bar{Q}$ can be expressed in
terms
of the multiplicity in light quark production as$^{[15,18]}$
\begin{equation}
\Delta N^{Q\bar{Q}}(W) = N^{q\bar{q}}(W) - N^{q\bar{q}}(\sqrt{e}M_Q) +
O(\alpha_s(M^2_Q)N(M_Q))
\end{equation}
with $\Delta N$ the accompanying hadron multiplicity.  An immediate consequence
is that the difference $\delta_Q$ between particle yields from $q$ and $Q$-jets
remains $W$ independent.  This QCD result contradicts the naive expectation
\begin{equation}
\Delta N^{Q\bar{Q}}(W) = N^{q\bar{q}} (W(1 - <x>_Q))
\end{equation}
according to which $\delta_Q$ would be a decreasing function of $W$.  The
existing data on $b$-quarks are consistent with the MLLA and the hypothesis (4)
appears to be disfavoured$^{[19]}$.  Further detailed measurements, especially
with $c$-quark events, could provide stringent tests of MLLA-LPHD predictions.
The study of $b$-quark events at the energies of LEP 2 looks rather promising.
In this region the naive formula (4) predicts a negative value of $\delta_b$
(about -(3 $\div$ 4)) in marked contrast with the LEP 1 result ($(\delta
_b)_{exp} \sim 3)$.

Finally, let us mention that LEP 2 could be a useful laboratory for further
studies of colour-related interjet phenomena, such as the celebrated
string/drag
effect in 3-jet events.

\vspace*{.5cm}

\noindent  2.2.  Hard QCD physics

Studies of hard QCD usually concern measurements of the jet production rates,
hadronic event shapes, energy correlations etc., see [3,4].  We shall
illustrate
here the potential of LEP 2 to examine the running of the strong coupling
constant basing on the 3-jet event production rate $R_3$.  The foreseen
luminosity of 500 pb$^{-1}$ would provide measurement of the ratio
$\frac{\alpha_s(LEP 2)}{\alpha_s(LEP 1)}$ with a statistical error of about
2\%$^{[3]}$.

A summary of measurements of $R_3$ is presented in Fig.\ 3 borrowed from [4].
\begin{figure}
\vspace*{10cm}
\end{figure}
As one can easily see, the high-order terms affect the energy dependence of
$R_3$ only slightly.  The existing data provide quite convincing evidence for
the logarithmic decrease of $\alpha_s$.  The significance will be much
increased
with the measurements at LEP 2.  Here the possibility to use the same detector
(in order to eliminate experimental point-to-point uncertainties) in a wide
energy range is especially beneficial.  I would like to stress that I do not
belong to the \lq\lq light gluino club".  The gluino curve in Fig.\ 3 aims only
to demonstrate the potential of LEP 2 and NLC for this type of study (see Ref.\
[4] for details and references).  It is important to achieve the precision
necessary to distinguish between the standard QCD and the scenarios like light
gluino hypothesis.

\vspace*{1cm}

\noindent  {\bf 3.  Hadronic $W^+W^-$ Events and Colour Rearrangement Effects}

QCD interference effects between  $W^+$ and $W^-$ undermine the traditional
meaning of a $W$ mass in the process $e^+e^- \rightarrow W^+W^- \rightarrow
q_1\bar{q}_2q_3\bar{q}_4$.  Specifically, it is not even in principle possible
to subdivide the final state into two groups of particles, one of which is
produced by the $q_1\bar{q}_2$ system of the $W^+$ decay and the other by the
$q_3\bar{q}_4$ system of the $W^-$ decay: some particles originate from the
joint action of the two systems.  Since a determination of the $W$ mass is one
of the main objectives of LEP 2, it is important to understand how large the
ambiguities can be.  A statistical error of 55 MeV per experiment is
expected$^{[1,2]}$, so the precision of the theoretical predictions should
match
or exceed this accuracy.  A complete description of interference effects is not
possible since non-perturbative QCD is not well understood.  The concept of
colour reconnection/rearrangement is therefore useful to quantify effects.  In
a
reconnection two original colour singlets (such as $q_1\bar{q}_2$ and
$q_3\bar{q}_4$) are transmuted into two new ones (such as $q_1\bar{q}_4$ and
$q_3\bar{q}_2$).  Subsequently each singlet system is assumed to hadronize
independently according to the standard algorithms.  Depending on whether a
reconnection has occurred or not, the hadronic final state is then going to be
somewhat different.  The reconnection effects were first studied in Ref.\ [20]
but these results were mainly qualitative and were not targeted on what might
actually be expected at LEP 2.  The picture in [20] represents an example of
the
so-called instantaneous reconnection scenario, where the alternative colour
singlets are immediately formed and allowed to radiate perturbative gluons.
For
a detailed understanding of QCD interference effects in $W^+W^-$ events one
needs to examine the space-time picture of the process.  A systematic analysis
of QCD rearrangement phenomena in $W^+W^-$ events has been performed in Ref.\
[21] (see also [22]).  It was shown that interference is negligibly small for
energetic perturbative gluon emission.  Firstly, the $W^+$ and $W^-$ decay at
separate times after production, which leads to large relative phases for
radiation off the two constituents of a rearranged system, and a corresponding
dampening of the QCD cascades$^{[23]}$.  Secondly, within the perturbative
scenario the colour transmutation appears only in order $\alpha^2_s$ and is
colour-suppressed.  It was concluded that only a few low-energy particles could
be affected.   In order to understand the reconnection effects occuring at the
non-perturbative hadronization stage, the standard Lund fragmentation
model$^{[24]}$ has been considerably extended and several alternative models
for
the space-time structure of the fragmentation process have been developed.
Comparing different models with the no-reconnection scenario, it turns out that
reconnection effects are very small.  The change in the averaged charged
multiplicity is at the level of a per cent or less, and similar statements hold
for rapidity distributions, thrust distributions and so on.  The total
contribution to the systematic error on the $W$ mass reconstruction may be as
large as 40 MeV.  This is good news.  Otherwise, LEP 2 would not have
significant advantages in the measurements of $M_W$ over hadronic machines
where
the accuracy is steadily improving$^{[25]}$.  Let us emphasize that in view of
the aimed-for precision, 40 MeV is non-negligible.  However, remember that as a
fraction of the $W$ mass itself it is a half a per mille error.  Reconnection
effects are therefore smaller in the $W$ mass than in many other observables,
such as the charged multiplicity etc.

Clearly, colour rearrangement effects are interesting in their own right, for
instance, as a new probe of the non-perturbative QCD dynamics$^{[20-22]}$.
However, the standard measures considered in [21] seem to be below the
experimental precision one may expect at LEP 2.  A more optimistic conclusion
has been reached in Ref.\ [22], where some specific ways to disentangle colour
reconnection phenomena were proposed.  But personally I still believe that one
will need good luck in order to establish the nature and size of the QCD
rearrangement effects in real-life experiments.

\vspace*{1cm}

\noindent  {\bf 4.  Gamma-gamma Topics}

One of the most promising areas for QCD tests at LEP 2 will be the study of
photon-photon collisions in the processes $e^+e^- \rightarrow e^+e^- +$
hadrons.
Two-photon physics is a remarkably rich subject, having elements in common with
both $e^+e^-$ and hadron-hadron collisions.  Photons can interact in different
ways: as vector mesons, as partons, or through their quark-gluon content.
There
has been a lot of recent activity in this field covering various aspects of
$\gamma\gamma$ physics\footnote{Recent reviews of theoretical and experimental
results are given in Refs.\ [26-29] and in the proceedings of the International
Workshops on photon-photon collisions.}.  I have not much to add to the
existing
comprehensive studies and I will only enumerate here a few topics where some
significant progress is expected from the measurements at LEP 2.
\begin{enumerate}
\item  Studies of the photon structure function $F^{\gamma}_2(x,Q^2)$

This classic probing of QCD in $\gamma\gamma$ collisions can be extended at LEP
2 to the higher regime of $Q^2$ up to $\sim 10^3$ GeV$^2$ (for some details see
Ref.\ [30]).  Further definite tests of the QCD prediction that $F^{\gamma}_2$
rises linearly with log$Q^2$ can be made here.  At the same time, one can go
down significantly further in $x$ ($x_{min} \geq 0.0015$ for $Q^2 = 15$
GeV$^2$).  One of the challenging tasks here is to find a rapid rise of
$F^{\gamma}_2$ at low $x$ analogous to that reported by experiments at HERA.
The measurements would provide important information on the low-$x$ dynamics
which is now a field of strong theoretical interest.
\item  One of the interesting $\gamma\gamma$ topics at LEP 2 is the exploration
of heavy quark production in the newly accessible region.  This process has
been
analysed in detail in Ref.\ [31].  The interest in this subject has been
boosted
by the new measurements from TRISTAN, indicating a possible excess of open
charm
production over the QCD calculations, see Ref.\ [27].
\item  The production of jets (and multi/mini jets)

Here the combination of different processes (VDM-like, direct, resolved)
provides one with the extra degrees of freedom in testing QCD theory.  This
subject by itself is a vast phenomenological field and it needs special
detailed
reviewing.  It is worth mentioning that especially in the case of the so-called
minijet production, the uncertainties in calculations are still quite
large$^{[32]}$.
\item  Traditional studies of $\gamma\gamma$ cross-sections and global event
properties, and their comparison with $pp$ and $\gamma p$ events. More activity
(transverse energy flow, multiplicity, jet rate ...) is predicted in
$\gamma\gamma$ events than in $\gamma p$ and $pp$ ones, see Ref.\ [28].  Fig.\
4
illustrates
\begin{figure}
\vspace*{10cm}
\end{figure}
the $\frac{dE_T}{dy}$ flow for $\gamma\gamma$ c.m.\ energy of 25 GeV.  Further
interesting issues can be addressed when either photon or both of them are
virtual.  This would open a quite new window on the structure of the photon.
\end{enumerate}

\vspace*{1cm}

\noindent  {\bf 5.  Summary}

This talk represents a (rather biased) attempt to highlight the potential of
LEP
2 for QCD studies.  The main conclusion can be stated as follows.   With a lot
of work and some luck, LEP 2 will still have much to teach us about QCD
physics.

\vspace*{1cm}

\noindent  {\bf Acknowledgements}

I wish to thank D.J.\ Miller, T.\ Sj\"{o}strand and W.J.\ Stirling for fruitful
discussions. This work was supported by the United Kingdom Science and
Engineering Research Council.

\newpage

\noindent  {\bf Figure Captions}

\vspace*{2cm}

\begin{itemize}
\item[Fig.\ 1] Energy evolution of the peak position compared to a prediction
of
Eq.\ (2).

\vspace*{3cm}

\item[Fig.\ 2] $\ell n \frac{1}{x_p}$ distribution of charged particles at $W$
=
30,90,200 and 500 GeV computed with the JETSET program (recall that the spectra
are practically blind/democratic towards the WIG'ged MC's).  The figure is
borrowed from the presentation of G.\ Cowan in [6].

\vspace*{3cm}

\item[Fig.\ 3] Energy dependence of $R_3$ (JADE algorithm, $y_{cut} = 0.08$)
compared with analytic QCD calculations and with the hypothesis of the
existence
of a light gluino.$^{[4]}$

\vspace*{3cm}

\item[Fig.\ 4] Transverse energy flow at $E_{cm} = 25$ GeV as a function of
rapidity for different beams.$^{[28]}$  $\gamma\gamma$: full histogram; $\gamma
p$: dashed one; and $pp$: dash-dotted one.

\end{itemize}

\end{document}